# Tuning and Switching a Plasmonic Quantum Dot 'Sandwich' in a Nematic Line Defect

*Haridas Mundoor[†], Ghadah. H. Sheetah[‡], Sungoh Park[†], Paul J. Ackerman[†], Ivan I. Smalyukh[\*†‡ ∥ ⊥] and Jao van de Lagemaat[\*† ⊥ §].*

[†] Department of Physics, University of Colorado, Boulder, CO 80309, USA.

[‡] Materials Science and Engineering Program, University of Colorado, Boulder, CO 80309, USA.

[∥] Soft Materials Research Center and Department of Electrical, Computer and Energy Engineering, University of Colorado, Boulder, CO 80309, USA.

[⊥] Renewable and Sustainable Energy Institute, National Renewable Energy Laboratory and University of Colorado, Boulder, CO 80309, USA.

[§] National Renewable Energy Laboratory, Golden, Colorado 80401, USA.

## ABSTRACT

We study the quantum-mechanical effects arising in a single semiconductor core/shell quantum dot controllably sandwiched between two plasmonic nanorods. Control over the position and the "sandwich" confinement structure is achieved by the use of a linear-trap, liquid-crystal line defect and laser tweezers that 'push' the sandwich together. This arrangement allows for the study of exciton-plasmon interactions in a single structure, unaltered by ensemble effects or the complexity of dielectric interfaces. We demonstrate the effect of plasmonic confinement on the photon-antibunching behavior of the quantum dot and its luminescence lifetime. The quantum dot behaves as a single emitter when nanorods are far away from the quantum dot but shows possible multiexciton emission and a significantly decreased lifetime when tightly confined in a plasmonic 'sandwich'. These findings demonstrate that liquid crystal defects, combined with laser tweezers, enable a versatile platform to study plasmonic coupling phenomena in a nanoscale laboratory, where all elements can be arranged almost at will.

The properties of nanoparticles, often dramatically different from those of bulk materials despite identical chemical composition, can be further controlled through interactions with other nanoparticles[1-3]. The exploration of plasmonic phenomena as a means to control the behavior of excitonic systems is a field of active study[4-15]. Such interactions can be used to detect single molecules[16,17], to drive up- and down-conversion[18,19], to enhance optical absorption in solar cells[20-22], to induce hot-electron charge transfer[23], to drive solar-to-fuel photocatalytic systems,[24] and more. There is a great need for the exploration of plasmon-exciton interactions at the level of individual nanoparticles and nanostructured composites, though such experimental explorations are often made challenging by the sensitivity of these interactions to the proximity of dielectric interfaces, which can cause various artifacts[7]. To mitigate these problems, we recently used laser tweezers and a point defect in the bulk of a liquid crystal (LC) to study the exciton/plasmon coupling of a co-entrapped quantum emitter and a single plasmonic particle.[7] This arrangement allowed for comparison of blinking characteristics of the quantum emitter before and after introducing a plasmonic particle to the LC point defect and showed that the radiative lifetime of the quantum dot (QD) decreased by an order of magnitude, resulting in a reduction of the blinking of the QD's luminescence and a concomitant increase in its radiative efficiency because of the Purcell effect. While this demonstrated the use of LC systems as nanoscale manipulation tools, the enhancement to the local electric field was induced with complex-shaped "nanoburst" gold nanoparticles around which the plasmonic field enhancement was very inhomogeneous. Although this arrangement of nanoburst and quantum emitter turned out to be fortuitous, greater control over the plasmonic field and the placement of the plasmonic particles and the QD particle is needed.

In this paper, we demonstrate a system in which we use a topological line defect in the LC that traps QDs and well-defined gold nanorods (GNRs) along two dimensions but allows for controlled movement along one dimension (along the defect line) (Figure 1). The GNRs can be moved along the defect line by using infrared laser tweezers and can be manipulated to form a linear "sandwich" structure, where the QD is localized between the tips of two nanorods, so that it resides in a well-defined optical field geometry. In this way, we can study the emission behavior of a single dot, as well as how the emission is influenced when the QD is sandwiched between GNRs inducing a well-defined plasmon-enhanced electromagnetic field. We show that the plasmon coupling induces multiphoton emission and discuss how our findings may have an impact on the ability to design and realize novel mesostructured composite materials with novel physical behavior arising from controlled plasmon-exciton interactions.

**RESULTS AND DISCUSSION**

**Elastic trapping of nanoparticles in LC line defect.** We used 40 x 65 nm GNRs (Figure 1), which were synthesized by following the seed-mediated method described in detail elsewhere[25-27] (See materials and methods). We also utilized commercially available CdSe/ZnS core-shell QDs (Ocean nanotech) shown in Figure 1e, which were selected for their emission peak at 620 nm. These QDs are characterized by an average diameter of 10 nm with a CdSe core and a thin outer shell of ZnS forming a core-shell type structure. The design of our experiments was aimed to control the optical coupling between QDs and GNRs. For this reason, we chose to use GNRs that exhibit a longitudinal surface plasmon resonance (SPR) peak at about 620 nm, matching the emission peak of QD particles (as measured when deposited on a glass substrate). The 10 nm silica shells of GNRs effectively make their SPR properties insensitive to the LC's dielectric and order parameter tensor structures around nanoparticles within the surrounding host medium. As designed, our estimates show that, because of the shells, variations of the director structure within the LC do not cause shifts of SPR peaks for more than 3 nm. The core-shell QD particles also provide a high quantum efficiency while remaining chemically stable under different experimental conditions used in our work. Very dilute colloidal co-dispersions of GNR and QD nanoparticles (with number densities of each estimated to be 0.01 $\mu m^{-3}$ or even smaller) were obtained through a series of solvent exchanges in a chiral nematic LC with cholesteric pitch ~ 12 µm, and infiltrated into a glass cell with a gap of ~ 10 µm, as described elsewhere[7] (See materials and methods).

Free energy minimization for the confined frustrated chiral nematic LC with the cholesteric pitch to cell thickness ratio of ~1.2 promotes the formation of various cholesteric fingers and topological solitons like skyrmions, torons, and hopfions[28], all embedded in a uniform homeotropic background with the far-field director **n**$_0$ along the normal to substrates, as prescribed by the boundary conditions. These localized director structures arise to embed twisted regions within the unwound **n**$_0$–background, which helps to locally alleviate the frustration associated with the incompatibility of perpendicular surface boundary conditions and the helicoidal twist tendency. Among these solitonic configurations are the cholesteric finger structures of the third kind depicted in Figure 1a, which are of interest for the present study. Within the translationally invariant director structure of this finger, the director field **n(r)** twists by $\pi$ from left to right side of the cross-section (Figure 1a), with the twist handedness matching that of the equilibrium chiral LC, as determined by the chirality of the molecular chiral additive. Singular line defects within the structure of these fingers, which terminate the bulk $\pi$-twist of **n(r)** near the confining substrates to again match the director to perpendicular boundary conditions, serve as linear topological traps for both the plasmonic and semiconductor nanoparticles (Figure 1c). In a cross-section orthogonal to the axes of the finger and the defect lines, the molecular orientations of LC molecules and **n** around the cores (depicted as red wires in Figure 1b) of the defect lines rotate by $\pi$ while tracing the Mobius-strip-like patterns, with these orientations becoming undefined within the core (Figure 1b). Thus, these cores have reduced scalar order parameter as compared to that of the bulk of LC and, in the simplest way, can be understood within the isotropic defect core model[29].

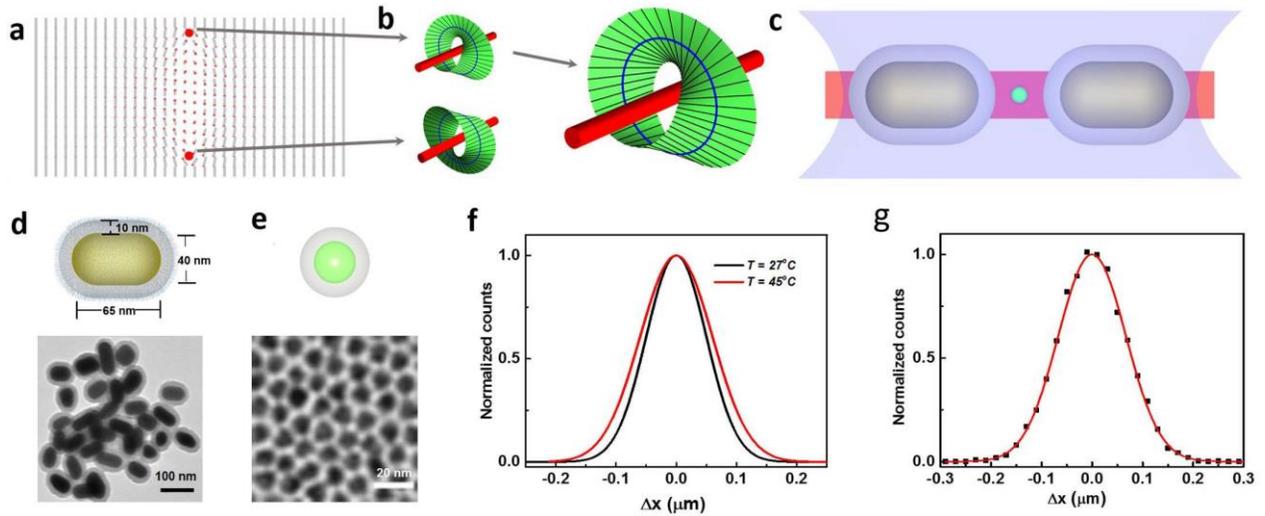

**Figure 1. Localization of GNR and QD nanoparticles in linear defect traps.** (a) Director configuration in the vertical cross-section of the cholesteric finger with two line defects (marked with the red filled circles), which is translationally invariant along the normal to the cross-section. (b) Configurations of the director field around the defect lines at the top and bottom of the cross-section shown in (a), with details of the director configuration for one of them shown on the right side. The core regions of defects are shown using red color. (c) A schematic illustration of the experimental configuration of QD (green) and GNR (yellow) nanoparticles co-entrapped within the core of a singular line defect (red tube) within a chiral nematic LC. (d) Schematics of the GNR particle with a silica shell and DMOAP surfactant monolayer (top) with dimensions marked on the illustrations and TEM micrograph of the silica capped GNRs (bottom). (e) Schematics of the CdSe/ZnS QD particle representing a core-shell geometry and TEM micrograph of the QD particle (bottom) used in the experiments. (f) The probability distribution of the displacement made by a single GNR within time periods $\Delta t = 0.067$ s, showing the diffusion along the length of the line defect at room temperature and at 45° C. (g) The probability distribution of the displacement made by a QD in time $\Delta t = 0.067$ s, describing its diffusion along the length of the line defect.

Because of the associated increased free energy density, singular defect core regions can act as linear traps to spatially confine various nanoparticles (Figure 1c)[30]. Confinement of nanoparticles within a defect core reduces the overall free energy by displacing the energetically costly defect region with the nanoparticle. The nanoparticles entrapped inside of these defect regions are strongly nano-confined in terms of their spatial positions in a plane orthogonal to the line defect (Figure 1c), but undergo Brownian motion and diffuse freely along the length of the line defect due to thermal energy (Figure 1f,g). The one-dimensional diffusion coefficients measured for individual defect-entrapped GNRs and QDs are estimated to be, respectively, $1.9 \times 10^{-2}$ μm$^2$s$^{-1}$ and $3.4 \times 10^{-2}$ μm$^2$s$^{-1}$ at room temperature, though they increase as the temperature is increased (for example, at the elevated temperature of 45°C, we obtain the diffusion coefficient $D_{GNR} = 2.7 \times 10^{-2}$

µm²s⁻¹ for the GNRs). Furthermore, the rod-like GNRs spontaneously align with their long axes along the defect line (Figure 1c) because this maximally reduces the free-energy-cost of the core by maximally displacing high-energy regions with the volume of the nanoparticle. Although a detailed quantitative picture of the interactions between a singular defect line and nanoparticles requires numerical modeling based on the tensorial order parameter[31], one can gain insights into the physical underpinnings behind our observations *via* simple estimates of potential energies involved. The simplest model of the defect line's singular core assumes that this core is an isotropic liquid and that its energetic cost is proportional to $k_B \Delta T_c$, where $k_B$ is the Boltzmann constant and $\Delta T_c = T_{NI} - T$ is the difference between the nematic-isotropic transition temperature $T_{NI}$ and the sample's absolute temperature $T$ [31]. The temperature dependence of diameter of the singular line's core can be estimated by comparing the free energy cost of melting LC to the isotropic state and the free energy of producing strong elastic distortions around the defect line's core[31], which yields $d_c = (2MK/(\rho N_A k_B \Delta T_c))^{1/2}$, where $K$ is the average Frank elastic constant, $M$ is the molecular mass, $N_A$ is the Avogadro's number and $\rho$ is the density of the LC. Although a typical value for $d_c$ deep in the nematic phase is ≈10 nm [32,33], by varying $\Delta T_c$ one can tune $d_c$ in the range 10-100nm. This capability provides a key advantage to the implementation of our experiments, as discussed below. The defect line's free energy reduction associated with the placement of a nanoparticle within its core can be estimated using the free energy cost per unit length of a half-integer disclination line $W \approx (\pi/4)K\ln(2R/d_c) + W_c$, where $R$ is a characteristic dimension of the sample and $W_c$ is energy per unit length of the isotropic disclination core[31]. By following Ref.[31], one can calculate the defect's energies reduction that enable nanoparticle trapping as ≈ $23 k_B T$ for QDs and ≈ $450 k_B T$ for GNRs. Laser tweezers at moderate laser powers of up to 50 mW that we use do not exert optical trapping potential that would be sufficient to trap QDs, consistent with the well-known limitations of optical trapping of objects that are ≈10nm or smaller in size[34], or to remove QDs or GNRs from the singular line traps[31], or to reorient GNRs away from the long axis parallel to the defect line[31], though they allow for the translation of GNRs along the defect lines while controllably forming the sandwich structure.

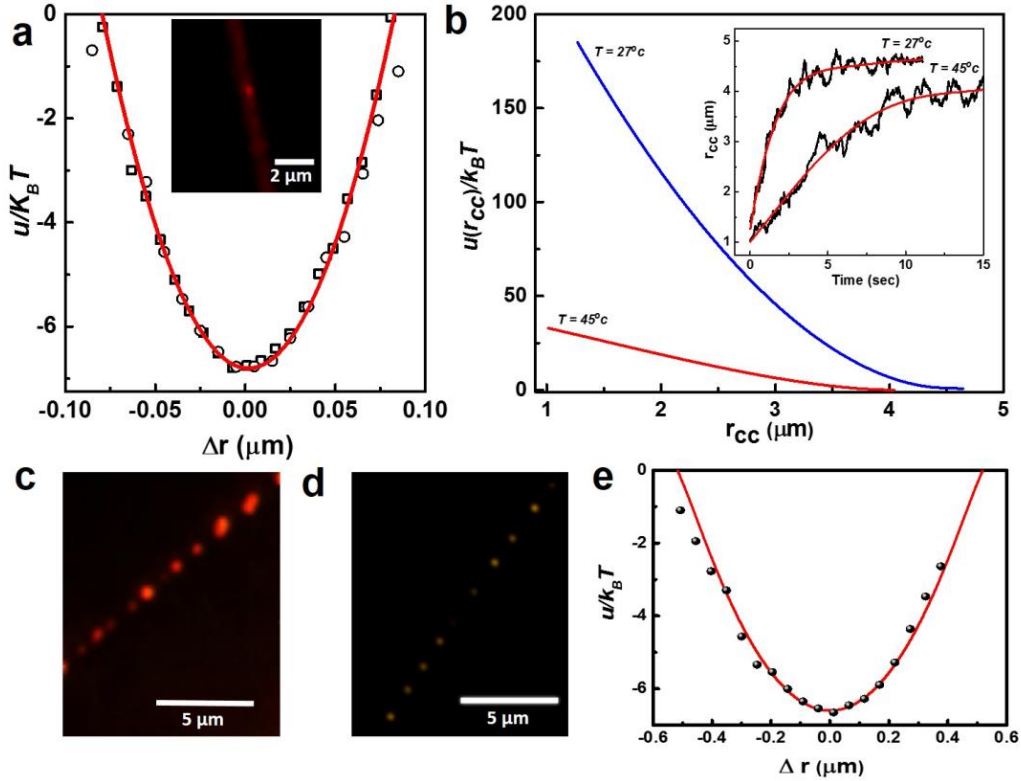

**Figure 2. Trapping and interactions of nanoparticles in the cores of defect lines.** (a) Trapping potential extracted from the motion of a QD (○) and a GNR (□) in directions perpendicular to the line defect. Inset shows the image of QD particles trapped inside line defect. (b) Interaction potential between two GNR entrapped inside a core of the line defect, probed when they are brought close to each other using optical trapping and then released. The potential is measured at room temperature and at 45 °C. Inset shows the variation of inter-particle center-to-center spacing with time when the optical traps are switched off, demonstrating repulsion between the nanoparticles. (c,d) Assemblies of (c) small clusters of QDs and (d) individual GNRs in a line defect formed due to repulsive interactions and confinement along the defect line. (e) Potential energy of interaction between GNRs extracted from the motion of single particles within an assembly shown in (d).

To quantify the strength of the topological defect line traps, we have characterized their stiffness. By measuring the small spatial displacements arising from the Brownian motion of the entrapped GNR and QD particles in the direction perpendicular to the defect line with the help of video microscopy (Figure 2a), we find the defect trap's stiffness $\xi_{DT}$= 4.2 pN/μm. This ensures a robust one-dimensional topological defect confinement of nanoparticles in our experiments. The studied silica-coated GNRs can be also effectively manipulated with the help of optical tweezers utilizing a 1064 nm fiber laser, which are described in detail elsewhere[35]. Once entrapped within a line defect's core, GNRs can be optically translated by dragging the tightly focused infrared laser beam of the tweezers along the length of the line defect, which helps to realize the desired ideal configuration of nanoparticles depicted in Figure 1c. Within this basic configuration used in our work (Figure 1c), a single QD nanoparticle is sandwiched between two GNRs, all positioned with their centers of mass, on average, along a single straight line coinciding with the central axis of the defect core. The GNRs are manipulated with laser tweezers so that the distance between them can be varied, as further detailed below, which is important for the exploration of plasmon-exciton interactions associated with these nanoparticles. Such a linear configuration of spatially co-localized nanoparticles prompts strong changes in the photophysics of the QD particle, owing to the interaction between the excitonic resonance of the QD and surface plasmon resonance of the GNR.

**Elastic interactions between the nanoparticles in a LC line defect.** GNR particles, along with the silica shells and surfactant monolayers around them, have a diameter in the range 50-55nm, larger than the effective diameter of the reduced-order core regions of the line defect. The DMOAP functionalization assures perpendicular surface boundary conditions for **n(r)** at their surface in the locations where parts of this surface protrude outside of the "melted" reduced-order defect core (Figure 1c). These radial (locally perpendicular to the cylindrical surface) boundary conditions for **n(r)** at the GNR periphery are incompatible with the π-twisted Mobius-strip-like **n(r)**-structure of the twist disclination around its core. This causes further elastic distortions, which effectively create a repulsive interaction between the GNR particles (the repulsion stems from the fact that these additional elastic distortions are squeezed to a smaller region when two GNRs are brought closer along the defect line, which costs more elastic energy). We have characterized the interaction potential between two GNR particles by tracing the inter-particle spacing with time when they are moved close to each other with the help of optical tweezers and then released (Figure 2b). When the temperature of the LC is increased, the effective diameter of the melted defect core becomes larger, leading to a reduction in the repulsive interaction potential, as demonstrated in Figure 2b. Due to the repulsive interaction, GNR particles entrapped inside the line core of the line defect form an assembly of equally spaced particles, with the one-dimensional crystal-like order emerging at sufficiently high concentrations and tweezer-assisted or other types of confinement along the length of the defect line (Figure 2c,d). The maximum repulsive potential ranges from 10 to 200 $k_BT$ (Figure 2b), making one-dimensional dispersions of GNRs along the defect cores robust with respect to thermal fluctuations (Figure 2e). Under these conditions, relatively large GNRs mutually repel and never aggregate, though they can be pushed towards each other to effectively form dimers at elevated temperatures (still below the nematic-isotropic transition temperature) with the help of laser tweezers, which we will utilize in the experiments described below.

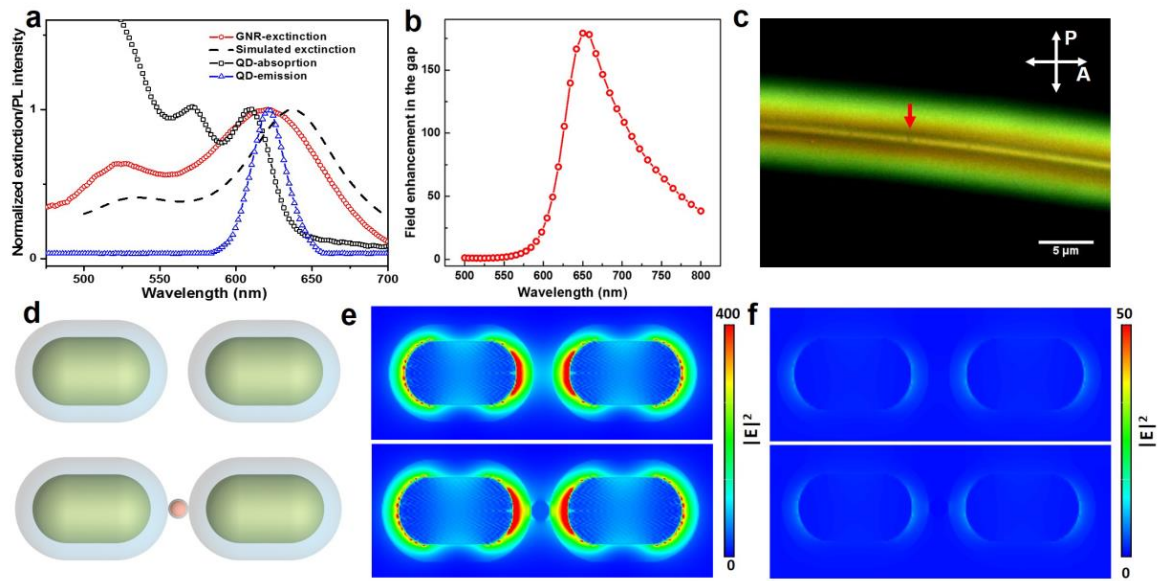

**Figure 3. Optical characterization of nanoparticles and modeling of SPR effects involving them.** (a) Optical characterization of the nanoparticles used in the experiments showing extinction spectra of GNR (○) particles dispersed in water, simulated extinction spectra of GNRs when they are brought close to each other in the line defect, forming a dimer configuration, indicating a red-shift in the longitudinal LSPR peak. Absorption spectra of the QDs (□) dispersed in toluene and emission spectra of the QDs particles on a glass substrate (Δ). (b) Electric field enhancement at the QD location in the sandwich structure for different emission wavelengths, calculated based on electromagnetic simulations using COMSOL Multiphysics. (c) Optical microscopy image of a LC line defect viewed under crossed-polarizers, indicating strong birefringence of the LC line defect. The location of a GNR- QD sandwich assembly is marked with a red arrow. (d) Schematic of the experimentally realized configuration of the particles showing the dimer configuration of the GNR particles with a QD particle located at the center of the GNR particles. (e,f) Electric field intensity profile for the configuration shown in (d) simulated using DDA method at emission wavelength (e) 620 nm and (f) excitation wavelength 473 nm.

Unlike GNRs, because their outer shell diameter is always smaller than the defect line core, QD particles entrapped inside line defects do not exhibit strong elasticity-mediated repulsive interactions. However, when the number density of QDs is high, they can interact with each other to form small aggregates, so that similar elasticity-mediated repulsive interaction between the aggregates emerge when they grow larger than the diameter of the defect core (Figure 2c). In the present study, however, we use vanishingly small number densities of QDs to ensure individually dispersed QDs entrapped within defect lines, which is further confirmed by characterization detailed below.

**Plasmon- exciton interaction studies.** In our experiments, we first locate a single QD inside a line defect on the basis of analyzing its luminescence blinking characteristics when the QD is excited with the 436 nm line from a mercury lamp. Individual GNR particles are detected/imaged using dark field microscopy, which can be done in parallel with their optical manipulation using laser tweezers. The experimental configuration depicted in Figure 1c is realized by using an optical tweezer to move the GNRs close to the QD particle in-between. To reduce the strength of repulsive interaction between the GNR particles, we increase the temperature of the LC medium with an objective heater (Bioptechs). The CdSe/ZnS QDs used in our experiments have an emission peak centered at 620 nm, matched to the longitudinal SPR peak of GNR particles, as shown in Figure 3a. Since the surface of the GNR is coated with a silica shell, the shift in the localized surface plasmon resonance (LSPR) peak due to the higher effective refractive index of the liquid crystal E7 is minimal. However, when two GNRs are brought close to each other, their longitudinal LSPR peak shifts to the higher wavelength owing to the plasmon coupling between the two GNRs[36,37](Figure 3a). Observation of the LSPR shift and measurements of the scattering spectra of the GNRs in our sandwich assembly are limited due to the strong birefringence (Figure 3c) and thermal fluctuations of the LC line defect. We estimate a redshift ~ 15 nm for the longitudinal LSPR peak of the GNR, when they are assembled to form a sandwich structure as depicted in Figure 3d (Supporting information Figure S3). It is evident that considerable spectral overlap between the QD emission and longitudinal LSPR of GNRs is present in the sandwich assembly. However, the $|E|^2$ field enhancement spectrum of the sandwich assembly peaks at wavelengths slightly longer than the absorption maximum as shown in Figure 3b and supporting information Figure S3.

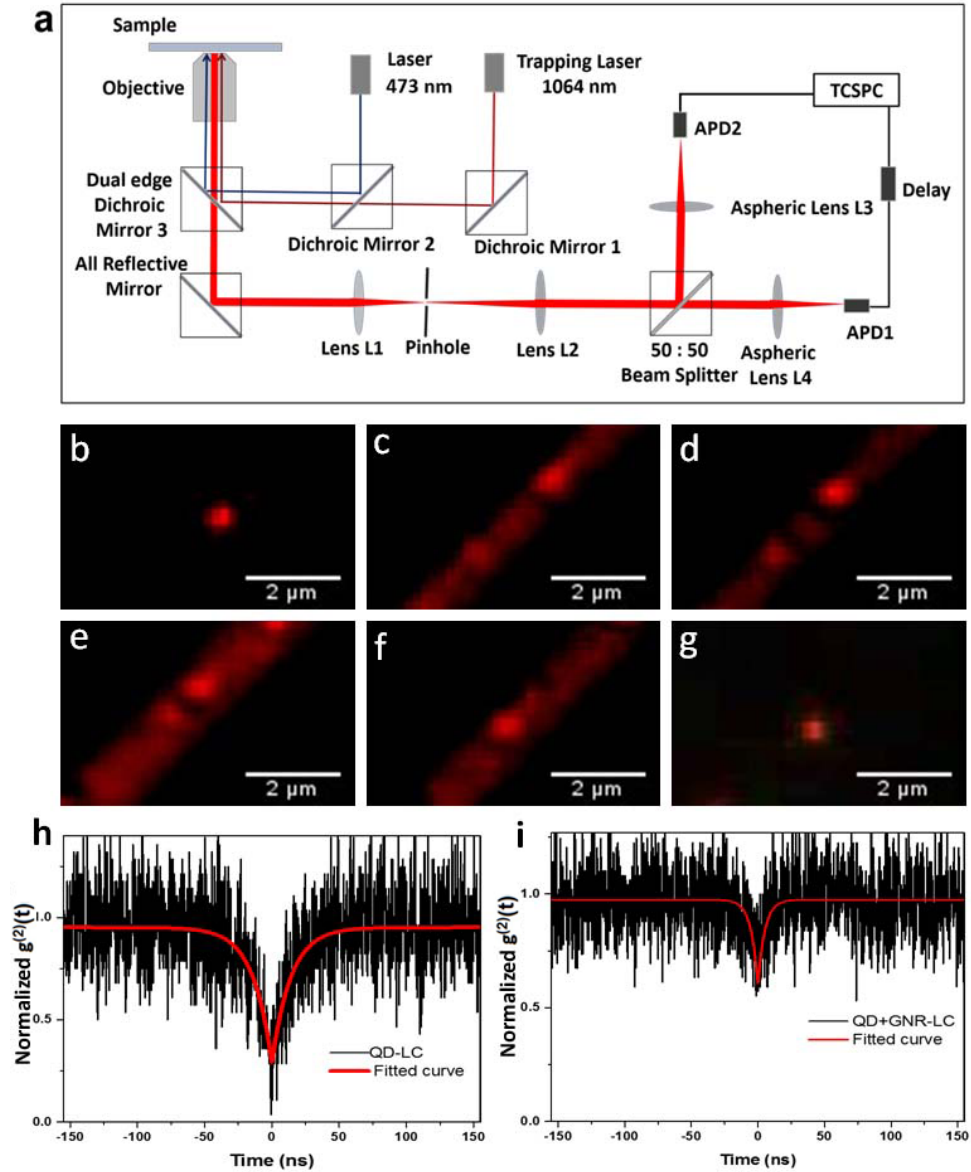

**Figure 4. Antibunching setup and characterization.** (a) Schematic representation of the antibunching setup used in experiment (b) Fluorescence image of a QD particle trapped inside a LC line defect before moving the GNR close to it. (c-f) Dark field microscopy images viewed with a red filter, showing the nanorod assembly using an optical tweezer, sandwiching a QD particle between. The final sandwich structure is represented in (f). (g) Fluorescence image of a QD particle after forming the sandwich structure. (h,i) Antibunching data collected from the QD particle before (h) and after (i) moving the GNR close to the QD, forming a sandwich assembly.

The linear arrangement of the GNRs, with the nanoscale separation between their surfaces controlled by tweezers, induces strong optical interaction between the SPR and QD emission due to the high Purcell factor achievable with such a configuration. To obtain additional insights into this phenomenon, we simulated the plasmonic field enhancement by using the Discrete Dipole Approximation (DDA) method[38-41]. We plot $|E|^2$ for an unpolarized light at wavelength 473 nm and 620 nm incident on the particles, as shown in Figure 3e, f. We verified these calculations by doing full electromagnetic simulations using COMSOL Multiphysics as shown in Figure. S3 (Supporting information). The strong plasmonic field enhancement (~ 250 at QD location) arising from the LPSR of GNRs at 620 nm is evident and mediates plasmon-exciton interactions that we discuss next.

We have analyzed the luminescence of individual QDs with the help of a photon antibunching setup (Figure 4a), which helps us to identify a single QD particle when it is far away from the GNRs. We first identify a single QD located inside the line defect as shown in Figure 4b, with two GNRs on both sides of the QD along the defect line but far away from it (Figure 4c). Light emission from this QD is analyzed using the antibunching technique, as shown in Figure 4h. The data is fit using the expression $g^{(2)}(t) = g^{(2)}(0) + (1 - e^{-\left|\frac{t}{\tau}\right|})/N$, where $\tau$ is the exciton lifetime, $N$ is the number of photons and $g^{(2)}(0)$ is the second order correlation at the coincidence time $t=0$[42,43]. Under the low excitation regime, the ratio of biexciton to exciton quantum yields of single QDs can be determined from the value of $g^{(2)}(0)$. Single photon emission is characterized by $g^{(2)}(0)= 0$ and $N=1$, although experimental observations are limited by the dark counts of the detector offsetting the value of $N$ and $g^{(2)}(0)$ as shown previously[44]. For the data shown in Figure 4h, we obtain $N =1.5$, $g^{(2)}(0)= 0.25$ indicating single photon emission, with an exciton lifetime $\tau = 14.2$ ns, which is close to the value previously reported for CdSe QDs with core-shell structure[45]. Following this characterization of the QD alone, we moved the GNR particles closer to the QD by using the optical tweezer, realizing the "sandwich" configuration depicted in Figure 1c. In order to reduce the repulsive interaction between the GNRs, the sample temperature was raised to 45°C with an objective heater and cooled to room temperature after the assembly process (Figure 4b-g).

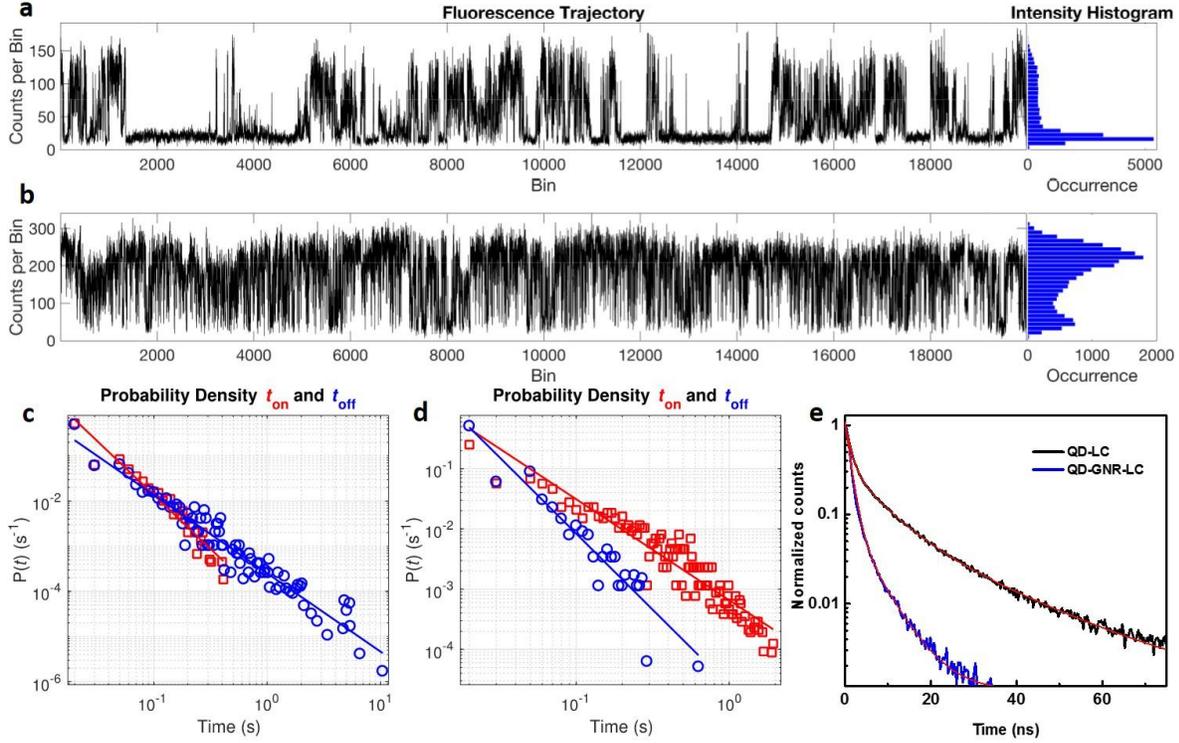

**Figure 5. Characterization of fluorescence intermittency and fluorescence decay**. (a, b) Fluorescence time traces of a QD particle trapped inside a line defect before (a) and after bringing two GNRs forming a sandwich assembly (b). The "on" and "off" times of the QD particle is presented by the corresponding histogram in the right side. (c, d) Analysis of fluorescence time trace with constant thresholding for the curves presented in (a) and (b) representing the probability density P(t) of sustained "on" ($t_{on}$) and "off" ($t_{off}$) times of the QD particles before (c) and after (d) bringing two GNRs forming a sandwich assembly. Solid lines represent linear fits to the data, showing a power law dependence. (e) Typical fluorescence decay curves of a QD particle (black curve), representing a faster fluorescence decay when the QD particle is sandwiched between two GNRs (blue curve). Solid lines represent the double exponential fit to the experimental decay data.

We observed considerable changes in the antibunching data curve when the QD particle was squeezed between the GNRs, as shown in Figure 4i. Fitting similarly to that described above yields $N = 2.67$, $g^{(2)}(0) = 0.6$ and a faster radiative decay rate of $\tau = 5$ ns. These results are possible indications of multiphoton emission from QDs with a faster decay rate modified by the Purcell effect. In order to analyze the fluorescence enhancement and blinking characteristics of the QD particles, we measured the fluorescence trajectories of the particle before and after forming the sandwich assembly. Figure 5 a and b represent typical blinking trajectories of the QD particles collected with a binning time of 10 ms, without GNRs (Figure 5a) and within sandwich structure (Figure 5b), respectively. It is clear that the QD particle stays predominantly in the "on" state after sandwiching the particle between two GNRs as shown by the histogram presented in the right-hand side of the Figure 5a,b. Moreover, the QD emission intensity is enhanced by almost two times indicating the enhanced detection of the QD after forming the sandwich assembly caused by preferential beaming of the emission light in a "donut" pattern around the axis of the assembly due to the antenna-like arrangement. Furthermore, we have performed a detailed analysis of the blinking trajectories by constant thresholding method and calculated the probability distribution of "on" ($t_{on}$) and "off" ($t_{off}$) times of the fluorescence emission as shown in Figure 5c,d. A power law dependence, typical of the QD emission with increased "on" state probability is evident from Figures 5c and d. Additional insight to the exciton-plasmon interaction is gained by analyzing the fluorescence decay curves shown in Figure 5e. The decay curves can be well fitted using a biexponential function, $I = A_1 e^{-t/\tau_1} + A_2 e^{-t/\tau_2}$, where $\tau_1$ and $\tau_2$ are the slow and fast lifetime component, $A_1$ and $A_2$ represent the corresponding amplitudes, respectively. Before the formation of the sandwich assembly, the QD fluorescence yields typical values of decay time $\tau_1 = 1.1$ ns and $\tau_2 = 19$ ns. On the other hand, when sandwiched between GNRs, the faster decay component is more prominent and the lifetime of the slower decay component is also reduced to 5 ns, indicating more than three times increase in the decay rate. We identify the slower decay component associated with the radiative decay as corresponding to the single exciton emission. Although the reported values for the lifetime of the biexciton emission is of the order of hundreds of picoseconds, we believe the faster decay component in our measurements represents the radiative decay corresponding to the biexciton emission. Accurate estimation of the biexciton decay time in our experiment is limited by the time resolution of the TCSPC hardware. The values of decay rates estimated by fitting double exponential curves (Figure 5e) match well with the values extracted from the antibunching data (Figure 4h,i). Both experiments convincingly show that the proximity of plasmonic particles increases the radiative decay rate of the QD by more than three times, indicating a strong coupling of the QD emission with the SPR of GNR particles. The observed optical properties of the QD particle in sandwich assemblies were highly reproducible and found to be independent of the excitation polarization. Moreover, the analysis of emission intensity of the QD particle revealed minimal effect on the excitation polarization; which is expected due to the relatively modest field enhancement at the excitation wavelength as demonstrated in Figure 3f. Measurements of the fluorescence spectra of QD particle in the sandwich assembly indicate increased fluctuations in the fluorescence peak position with respect to the spectra collected before forming the sandwich assembly (Figure S2, Supporting information). The experiments were repeated on several QDs – GNR pairs in line defects forming a sandwich assembly and the observed optical properties are found to be highly reproducible (Figure 6). The values of $g^{(2)}(0)$, emission enhancement and fluorescence decay curves are shown in Figure 6a,b and supporting information Figure S1, Table S1.

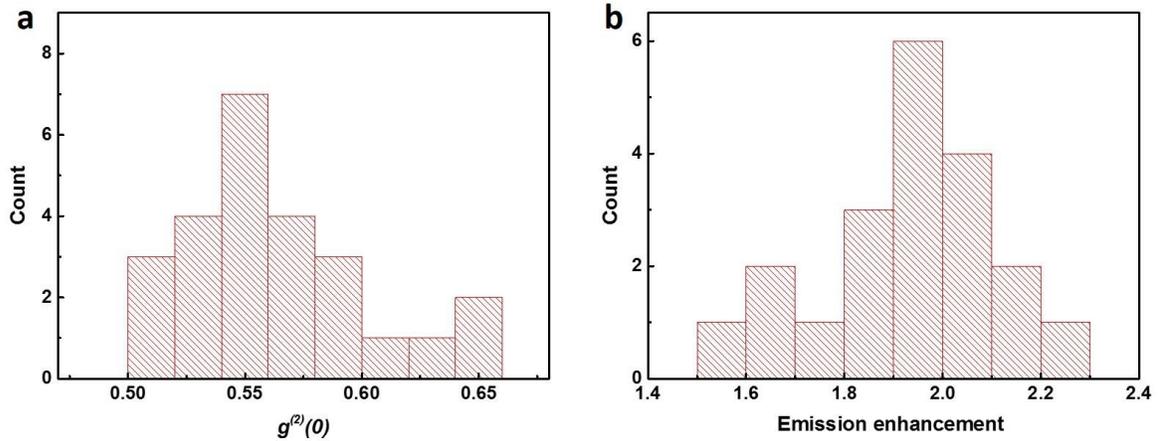

**Figure 6. Variations of $g^{(2)}(0)$ and enhancement factor.** Histogram representing the variations of $g^{(2)}(0)$ estimated from the photon antibunching measurements (a) and emission enhancement of QD fluorescence (b), calculated based measurements on multiple sandwich assemblies in LC line defect.

The study of multiexciton generation in semiconducting nanostructures has been intensively pursued over the last couple of decades[46,47]. Recent efforts in this field have been directed towards employing the LSPR of metal nanoparticles for generating multiexcitons in semiconducting nanoparticles[8-13], though studies on self-assembled nanoparticles have been rarely reported. Reports on QDs deposited on a roughened, gold-coated surface indicate enhanced multiphoton emission and decreased single photon emission on these substrates[8]. The quantum efficiencies of the multiexciton states in a QD are relatively low, due to Auger and other non-radiative decay channels quenching the multiphoton radiative process. The surface plasmon effects on the higher order exciton states are much different from the effects on the single exciton state. Significant enhancement of the radiative decay rates of the biexcitonic states can be achieved when coupled with SPR, effectively increasing the quantum efficiency of the biexciton states. In our experiments, emission from the QDs couples strongly with SPR of the GNR particles, which could enhance the biexcitonic radiative rates according to the Purcell effect, resulting in an increased probability of multiphoton emission from the QDs. At the same time, the radiative decay rate of the single exciton states are enhanced by almost 4 times, as revealed by the fluorescence decay measurements (Figure 5e) and also consistent with previous observations[7-13]. However, based on the fluorescence trajectories of the QD particles, collected before and after bringing the GNRs, the fluorescence intensity enhancement for the QD particle in the sandwich assemblies is only about 2 times which is expected due to the only modest enhancement of the light collection efficiency in the sandwich geometry. Although the nonradiative energy transfer between the QD and GNR particles can affect the emission properties of the particles, previous studies have shown that biexcitonic states are less affected by this interaction in comparison with the excitonic emission[10]. Moreover, as intended by our experimental design, specially-designed silica shells around the GNRs ensure that the QDs are separated from the GNR surface by about 10 nm, limiting the non-radiative energy transfer towards GNRs, preventing the complete quenching of the QD emission. In effect, coupling the SPR of the GNRs to the biexciton states of the QD allows for the branching of the biexciton state to multiphoton emission through the enhanced radiative rate of the biexcitonic states of QDs without significant change in the nonradiative Auger dominated recombination rate, effectively making the non-radiative decay channels of the biexcitonic states less prevalent. The excitation intensity aided by the field enhancement by the GNR provides sufficient field intensity for the generation of multiexciton states in the QD, even though the field enhancement at the excitation wavelength is relatively modest compared to the emission wavelength at 620 nm. The exact modelling of the SPR effect on the higher order states of the QD is challenging due to the complexities of the optical interactions involved. For example, plasmon modes give rise to a strong electric field gradient near the metal particle in the near field which could influence the dipole-forbidden, higher-order interactions in the QD emission[14,15]. A future successful model should include the interaction of higher-order exciton states with hot electrons, thermal and Auger relaxation in addition to accounting for the delocalization of electron and hole wave functions in case of core-shell type particles. Moreover, the interaction of semiconductor nanoparticles with surface ligands and the host medium will require separate detailed theoretical and experimental studies for more quantitative modeling of the role of SPR on the multiphoton emission from QDs.

## **CONCLUSIONS**

In conclusion, we have demonstrated the use of LC line defects to control the position and orientation of GNR and QD particles to study the optical interaction with a QD in a well-defined geometry and in the bulk of a dielectric LC medium, unaffected by the proximity to a substrate. Our results indicate possible multiphoton emission from a single QD particle when located in between two GNRs, which originates from the strong coupling with the SPR of these rod-like plasmonic nanoparticles. Although the present study was performed using individual nanoparticles, it provides insight into the possibility of designing mesostructured composite materials with arrays of defects entrapping both plasmonic and semiconductor nanoparticles, where optical properties of the medium can be potentially controlled by tuning the inter-particle spacing through the LC's facile response to external stimuli.

## **MATERIALS AND METHODS**

**GNR Synthesis.** GNRs used in our experiments were synthesized following the seed-mediated method described in detail elsewhere[25,26]. To prepare the seed, equal amounts of 5 mL of hexadecyltrimethylammonium bromide (CTAB, Sigma-Aldrich, 0.2M) and gold(III) chloride trihydrate ($HAuCl_4 \cdot 3H_2O$, Sigma-Aldrich, 0.5 mM) were added to a clean glass bottle followed by 0.6 mL of freshly prepared ice-cold sodium borohydride ($NaBH_4$, Sigma-Aldrich, 10 mM) under vigorous stirring for 2 minutes, and kept at 30ºC for 30 minutes before use, allowing the reaction to complete and $NaBH_4$ to fully decompose. In the meantime, the growth solution was prepared by mixing 25 µL silver nitrate (AgNO$_3$, Sigma-Aldrich, 16 mM) with 10 mL of CTAB (0.2 M), 10 mL of HAuCl$_4$ (1 mM) and 75 µL of L-ascorbic acid (Sigma-Aldrich, 80 mM) under vigorous stirring for 30 seconds. The growth process was initiated by adding 12 µL of seed solution into the growth solution, and then it was left undisturbed for 8 hours. The GNRs are separated from the solution by centrifugation at 9000 rpm for 30 minutes. These GNRs (Figure 1d) were coated with silica shells, which were tuned to be ~ 10 nm thick, and then surface-functionalized with dimethyloctadecyl[3-(trimethoxysilyl)propyl] ammonium chloride (DMOAP, Acros Organics) according to procedures described in our previous study[27]

**LC cell preparation.** We used a chiral nematic LCs by using a commercially available nematic mixture E7 (Slichem, China) doped with a chiral dopant Cholesteryl Pelargonate (Sigma-Aldrich). The concentration of the chiral additive in the mixture was adjusted to yield a desired cholesteric pitch ~12 µm [48]. LC cells used for the experiments were prepared by assembling two glass plates with a cell gap spacing of ~ 10 µm set using glass fibre segments mixed in a UV-curable glue. To obtain strong homeotropic boundary conditions for the orientation of the rod-like LC molecules and the director **n** describing their local average orientation, the inner surfaces of the glass plates were treated with DMOAP. Very dilute co dispersion of GNR and QD in LC were prepared by mixing the nanoparticle dispersions with LC followed by evaporation at 70°C. The LC mixture was infiltrated into the glass cells using capilary forces and sealed with a fast-setting epoxy.

**Experimental setup.** The experimental set up was built around an inverted microscope (IX 81, from Olympus), equipped with lasers, spectrometer and a CCD camera (Flea, PointGrey) for imaging. The microscope is integrated with a holographic optical tweezer operating at 1064 nm output from a fiber laser (IPG photonics). The excitation beam is sent to the microscope with the help of dichroic mirrors (Thorlabs DMSP805R, Semrock FF495-Di03-25x36) (Figure 4a) and focused onto the sample using a 100x oil immersion objective (Olympus, UPlanSApo, NA = 1.42). The emission from the QD particles is collected using the same objective and sent through the optical filters (Chroma HQ 610/75M) and pinhole before being analyzed by a Avalanche Photodiode (APD, from Pico Quant) or Hanbury Brown-Twiss interferometer arrangement for the antibunching analysis. The antibunching measurements were performed using the 473 nm CW output from a diode laser as the excitation source. The fluorescence signal incident on a 50:50 beam splitter is detected by two APDs, placed at the two out ports of the beam, as shown in Figure 4a. The electrical pulses generated by the APDs are analyzed using time-correlated single photon counting (TCSPC, SPC 130, from Becker and Heckle) hardware, by measuring the second-order cross-correlation between photons detected by the APDs with respect to the photon co-incidence time *t*. In order to obtain the correlation at the negative coincident time, the signal from one of the APD is delayed by 500 ns by employing delay electronics before being fed to the TCSPC. The fluorescence trajectories of the QD emission was recorded using an APD connected to a data acquisition board (NIDAQ-6363, National Instruments) and analyzed using a MATLAB code. For fluorescence decay measurements, 467 nm pulsed output from a diode laser (Nano LED- Horiba

Scientific, 1MHz, 200 ps) was used as the excitation source. The emission from the QD particles are detected using an APD and analyzed with the help of TCSPC hardware. The fluorescence spectra of a QD particle located in a LC line defect was measured using a spectrometer (SpectraPro-275, Acton Research Corporation) equipped with a grating of 600 g/mm. The detected spectra were recorded using an electron multiplying charge coupled device (EMCCD, iXon3 888, Andor Technology).

**Numerical modeling.** The electromagnetic field enhancement and LSPR peak of GNR in the sandwich assembly was simulated using the nanoDDSCAT$^+$ tool based on the DDA method[38-41]. For the DDA simulations, the structures of GNR and QD are first defined using a 3D computer graphics program, Blender v:2.75 (Figure 3d) and rendered to nanoDDSCAT$^+$ to form set of dipoles with refractive index values representing the gold core, silica shell and QD using a DDSCAT conversion tool[41]. We then calculate the extinction spectra and electric field intensity $|E|^2$ configuration around the GNRs in the sandwich structure. We have also estimated the electric field enhancement of GNRs in the sandwich structure using COMSOL Multiphysics. The incident electromagnetic wave excites the computational volume, enclosing the GNRs and QD. The simulation is performed in a spherical volume enclosed in a perfectly matched layer shell in the radial directions allowing the calculation of both scattering and absorption contributions.

**Supporting Information**

Fluorescence decay curve of QD particle in sandwich assembly and table of fitted lifetime values based on measurements on multiple QD assemblies. Fluorescence spectra of a QD particle in sandwich assembly, simulated extinction spectra of GNR in sandwich assembly, and calculated values of LSPR, maximum field enhancement wavelength with end-to-end separation between GNRs in LC line defect.

**Corresponding Author**

* Email: ivan.smalyukh@colorado.edu, jao.vandelagemaat@nrel.gov

**Author Contributions**

H.M, S.P, P.J.A, I.I.S and J.v.d.L performed experimental work and analyzed results. G.S synthesized and surface functionalized gold nanoparticles. S.P. and H.M did numerical modeling of the electric field enhancement due to GNRs in LC. All authors contributed to writing the manuscript. J.v.d.L and I.I.S. conceived and designed the project.

**Funding Sources**

We acknowledge the support of the Division of Chemical Sciences, Geosciences, and Biosciences, Office of Basic Energy Sciences of the US Department of Energy under Contract No. DE-AC36-08GO28308 with the National Renewable Energy Laboratory (S.P, J.v.d.L, and I.I.S) and also partial support of the NSF grant DMR-1410735 (H.M. and I.I.S.)

ACKNOWLEDGMENT

The authors thank P. Davidson, T. Lee and B. Senyuk for discussions and M. Almansouri for help with sample preparation. We acknowledge the use of nanoDDSCAT$^+$ tool available at nanohub.org, for the simulation of electric field enhancement in the QD-GNR sandwich confined in a line defect in the LC medium.

REFERENCES

1. Dey, S.; Zhao, J. Plasmonic Effect on Exciton and Multiexciton Emission of Single Quantum Dots. *J. Phys. Chem. Lett.*, **2016**, *7*, 2921–2929.
2. Livache, C. *et al*. Charge Dynamics and Optolectronic Properties in HgTe Colloidal Quantum Wells. *Nano Lett.* **2017**, *17*, 4067-4074.
3. Jiang, L.; Mundoor, H.; Liu, Q.; Smalyukh, I. I. Electric Switching of Fluorescence Decay in Gold–Silica–Dye Nematic Nanocolloids Mediated by Surface Plasmons. *ACS Nano* **2016**, *10*, 7064–7072.
4. Johnson, J. C.; Reilly, T. H.; Kanarr, A. C.; van de Lagemaat, J. The Ultrafast Photophysics of Pentacene Coupled to Surface Plasmon Active Nanohole Films. *J. Phys. Chem. C* **2009**, *113*, 6871–6877.
5. Ji, B.; Giovanelli, E.; Habert, B.; Spinicelli, P.; Nasilowski, M.; Xu, X.; Lequeux, N.; Hugonin, J.; Marquier, F.; Greffet, J.; Dubertret, B. Non-blinking quantum dot with a plasmonic nanoshell resonator. *Nat. Nanotechnol*. **2015**, *10*, 170–175.
6. Gómez, D. E.; Vernon, K. C.; Mulvaney, P.; Davis, T. J. Surface Plasmon Mediated Strong Exciton−Photon Coupling in Semiconductor Nanocrystals. *Nano Lett*. **2010**, *10*, 274–278.
7. Ackerman, P. J.; Mundoor, H.; Smalyukh. I. I.; van de Lagemaat, J. Plasmon–Exciton Interactions Probed Using Spatial Coentrapment of Nanoparticles by Topological Singularities. *ACS Nano* **2015**, *9*, 12392-12400.
8. LeBlanc, S. J; McClanahan, M. R; Jones, M; Moyer, P. J. Enhancement of Multiphoton Emission from Single CdSe Quantum Dots Coupled to Gold Films. *Nano Lett*. **2013**, *13*, 1662−1669.
9. Masuo, S.; Kanetaka, K.; Sato, R.; Teranishi, T. Direct Observation of Multiphoton Emission Enhancement from a Single Quantum Dot Using AFM Manipulation of a Cubic Gold Nanoparticle. *ACS Photonics* **2016**, *3*, 109−116.
10. Park, Y.-S.; Ghosh, Y.; Chen, Y.; Piryatinski, A.; Xu, P.; Mack, N. H.; Wang, H.-L.; Klimov, V. I.; Hollingsworth, J. A.; Htoon, H. Super-Poissonian Statistics of Photon Emission from Single CdSe-CdS Core-Shell Nanocrystals Coupled to Metal Nanostructures. *Phys. Rev. Lett*. **2013**, *110*, 117401.

11. Dey, S.; Zhou, Y.; Tian, X.; Jenkins, J. A.; Chen, O.; Zou, S.; Zhao, J. An experimental and theoretical mechanistic study of biexciton quantum yield enhancement in single quantum dots near gold nanoparticles , *Nanoscale* **2015**, *7*, 6851-6858.
12. Naiki, H.; Masuo, S.; Machida, S.; Itaya, A. Single-Photon Emission Behavior of Isolated CdSe/ZnS Quantum Dots Interacting with the Localized Surface Plasmon Resonance of Silver Nanoparticles. *J. Phys. Chem. C*. **2011**, *115*, 23299-23304.
13. Matsuzaki, K.; Vassant, S.; Liu, H. W.; Dutschke, A.; Hoffmann, B.; Chen, X.; Christiansen, S.; Buck, M. R.; Hollingsworth, J. A.; Götzinger, S.; Sandoghdar, V. Strong plasmonic enhancement of biexciton emission: controlled coupling of a single quantum dot to a gold nanocone antenna. *Sci. Rep*. **2017**, *7*, 42307.
14. Jain, P. K.; Ghosh, D.; Baer, R.; Rabani, E.; Alivisatos, P. A. Near-field manipulation of spectroscopic selection rules on the nanoscale. *Proc. Natl. Acad. Sci. U.S.A*. **2012**, *109*, 8016-8019.
15. Andersen, M. L.; Stobbe, S.; Sorensen, A. S.; Lodahl, P. Strongly modified plasmon–matter interaction with mesoscopic quantum emitters. *Nature. Phys*. **2011,** *7* **,** 215-218.
16. Zijlstra, P.; Paulo, P. M. R.; Orrit, M. Optical detection of single non-absorbing molecules using the surface plasmon resonance of a gold nanorod. *Nat. Nanotechnol*. **2012**, *7*, 379–382.
17. Kravets, V. G.; Schedin, F.; Jalil, R.; Britnell, L.; Gorbachev, R. V.; Ansell, D.; Thackray, B.; Novoselov, K. S.; Geim, A. K.; Kabashin, A. V.; Grigorenko, A. N. Singular phase nano-optics in plasmonic metamaterials for label-free single-molecule detection. *Nat. Mater*. **2013,** *12*, 304–309.
18. Rogobete, L.; Kaminski, F.; Agio, M.; Sandoghdar, V. Design of plasmonic nanoantennae for enhancing spontaneous emission. *Opt. Lett*. **2007**, *32*, 1623-1625.
19. Sun, Qi-C.; Mundoor, H.; Ribot, J. C.; Singh, V.; Smalyukh, I. I.; Nagpal, P. Plasmon-Enhanced Energy Transfer for Improved Upconversion of Infrared Radiation in Doped-Lanthanide Nanocrystals. *Nano Lett*. **2014**, *14*, 101−106.
20. Erwin, W. R.; Zarick, H. F.; Talbert, E. M.; Bardhan, R. Light trapping in mesoporous solar cells with plasmonic nanostructures. *Energy Environ. Sci*., **2016**, *9*, 1577-1601.
21. Atwater, H. A.; Polman, A. Plasmonics for improved photovoltaic devices. *Nat. Mater*. **2010**, *9*, 205–213.
22. Morfa, A. J.; Rowlen, K. L.; Reilly, T. H.; Romero, M. J.; van de Lagemaat, J. Plasmon-enhanced solar energy conversion in organic bulk heterojunction photovoltaics. *Appl. Phys. Lett*. **2008**, *92*, 013504.
23. Wu, K.; Chen, J.; McBride, J. R.; Lian, T. Charge transfer. Efficient hot-electron transfer by a plasmon-induced interfacial charge-transfer transition. *Science* **2015**, *349*, 632–635.
24. Thomann, I.; Pinaud, B. A.; Chen, Z.; Clemens, B. M.; Jaramillo, T. F.; Brongersma, M. L. Plasmon Enhanced Solar-to-Fuel Energy Conversion.*Nano Lett.*, **2011**, *11*, 3440–3446.

25. Perez-Juste, J.; Liz-Marzan, L. M.; Carnie, S.; Chan, D. Y. C.; Mulvaney, P. Electric-Field-Directed Growth of Gold Nanorods in Aqueous Surfactant Solutions. *Adv. Funct. Mater.* **2004**, *14*, 571.
26. Ye, X.; Jin, L.; Caglayan, H.; Chen, J.; Xing, G.; Zheng, C.; Doan-Nguyen, V.; Kang, Y.; Engheta, N.; Kagan, C. R.; Murray, C. B. Improved size-tunable synthesis of monodisperse gold nanorods through the use of aromatic additives. *ACS Nano* **2012**, *6*, 2804.
27. Sheetah, G. H.; Liu, Q.; Smalyukh, I. I. Self-assembly of predesigned optical materials in nematic codispersions of plasmonic nanorods. *Optics Letters* **2016**, *41*, 4899.
28. Ackerman, P. J.; Smalyukh, I. I. Diversity of Knot Solitons in Liquid Crystals Manifested by Linking of Preimages in Torons and Hopfions. *Phys Rev X* **2017**, *7*, 011006.
29. Chaikin, P. M.; Lubensky, T. C., *Principles of Condensed Matter Physics*, Cambridge Univ. Press, 1995.
30. Senyuk, B.; Evans, J. S.; Ackerman, P.; Lee, T.; Manna, P.; Vigderman, L.; Zubarev, E. R.; van de Lagemaat, J.; Smalyukh. I. I. Shape-Dependent Oriented Trapping and Scaffolding of Plasmonic Nanoparticles by Topological Defects for Self-Assembly of Colloidal Dimers in Liquid Crystals. *Nano Lett* **2012**, *12*, 527-1114.
31. de Gennes, P. G.; Prost, J., *The Physics of Liquid Crystals,* 2nd ed. Clarendon, 1993.
32. Wang, X.; Miller, D. S.; Bukusoglu, E.; de Pablo, J. J.; Abbott, N. L. Topological defects in liquid crystals as templates for molecular self-assembly. *Nature Mater.* **2016,** *15*, 106–112.
33. Ravnik, M.; Zumer, S. Landau-de Gennes modelling of nematic liquid crystal colloids. *Liq. Cryst.* **2009**, *36*, 1201–1214.
34. Ashkin, A.; Dziedzic, J. M.; Bjorkholm, J. E.; Chu, S. Observation of a single-beam gradient force optical trap for dielectric particles. *Opt. Lett.* **1986**, *11*, 288-290.
35. Evans, J. S.; Sun, Y.; Senyuk, B.; Keller, P.; Pergamenshchik, V. M.; Lee, T.; Smalyukh, I. I. Active Shape-Morphing Elastomeric Colloids in Short-Pitch Cholesteric Liquid Crystals. *Phys. Rev. Lett.* **2013**, *110*, 187802.
36. Thomas, K. G.; Barazzouk, S.; Ipe, B. I.; Joseph, S. T. S.; Kamat, P. V. Uniaxial plasmon coupling through longitudinal self-assembly of gold nanorods. *J. Phys. Chem. B* **2004**, *108*, 13066-13068.
37. Jain, P. K.; Eustis, S.; El-Sayed, M. A. Plasmon coupling in nanorod assemblies: Optical absorption, discrete dipole approximation simulation, and exciton-coupling model. *J. Phys. Chem. B* **2006,** *110,* 18243-18253.
38. Draine, B. T.; Flatau, P. J. Discrete-dipole approximation for scattering calculations. *J. Opt. Soc. Am. A*, **1994** *11*, 1491-1499
39. Draine, B. T.; Flatau, P. J. Discrete-dipole approximation for periodic targets: theory and tests. *J. Opt. Soc. Am. A* **2008**, *25*, 2593-2703.
40. Flatau, P.J.; Draine, B. T. Fast near field calculations in the discrete dipole approximation for regular rectilinear grids. *Optics Express* **2012**, *20*, 1247-1252.

41. Jain, P.K.; Sobh, N.; Smith, J.; Sobh, A. R. N.; White, S.; Faucheaux, J.; Feser, J., **2015**, "nanoDDSCAT," https://nanohub.org/resources/dda. (DOI: 10.4231/D32V2CB3M).
42. Kitson, S. C.; Jonsson, P.; Rarity, J. G.; Tapster, P. R. Intensity fluctuation spectroscopy of small numbers of dye molecules in a microcavity. *Phys. Rev. A* **1998**, *58*, 620.
43. Hollars, C. W.; Lane, S. M.; Huser, T. Controlled non-classical photon emission from single conjugated polymer molecules. *Chem. Phys. Lett.* **2003**, *370*, 393–398.
44. Santori, C.; Pelton, M.; Solomon, G.; Dale, Y.; Yamamoto, Y. Triggered Single Photons from a Quantum Dot. Phys. *Phys. Rev. Lett*. **2001**, *86*, 1502.
45. Omogo, B.; Aldana, J. F.; Heyes, C. D. Radiative and Non-Radiative Lifetime Engineering of Quantum Dots in Multiple Solvents by Surface Atom Stoichiometry and Ligands. *J. Phys. Chem. C* **2013**, *117*, 2317.
46. Patton, B.; Langbein, W.; Woggon, U. Trion, biexciton, and exciton dynamics in single self-assembled CdSe quantum dots. *Phys. Rev. B* **2003**, *68*, 125316.
47. Park, Y. –S.; Malko, A. V.; Vela, J.; Chen, Y.; Ghosh, Y.; Garcia-Santamaria, F.; Hollingsworth, J. A.; Klimov, V. I.; Htoon, H. Near-Unity Quantum Yields of Biexciton Emission from CdSe/CdS Nanocrystals Measured Using Single-Particle Spectroscopy. *Phys. Rev. Lett.* **2011**, *106*, 187401.
48. Varney, M. C. M.; Zhang, Q; Senyuk, B; Smalyukh, I. I. Self-assembly of colloidal particles in deformation landscapes of electrically driven layer undulations in cholesteric liquid crystals. *Phys. Rev. E* **2016**, *94*, 042709.

# Supporting Information

# Tuning and Switching a Plasmonic Quantum Dot 'Sandwich' in a Nematic Line Defect

*Haridas Mundoor[†], Ghadah. H. Sheetah[‡], Sungoh Park[†], Paul J. Ackerman[†], Ivan I. Smalyukh[\*,†,‡,∥,⊥] and Jao van de Lagemaat[\*\*,†,⊥,§].*

[†] Department of Physics, University of Colorado, Boulder, CO 80309, USA.
[‡] Materials Science and Engineering Program, University of Colorado, Boulder, CO 80309, USA.
[∥] Soft Materials Research Center and Department of Electrical, Computer and Energy Engineering, University of Colorado, Boulder, CO 80309, USA.
[⊥] Renewable and Sustainable Energy Institute, National Renewable Energy Laboratory and University of Colorado, Boulder, CO 80309, USA.
[§] National Renewable Energy Laboratory, Golden, Colorado 80401, USA.

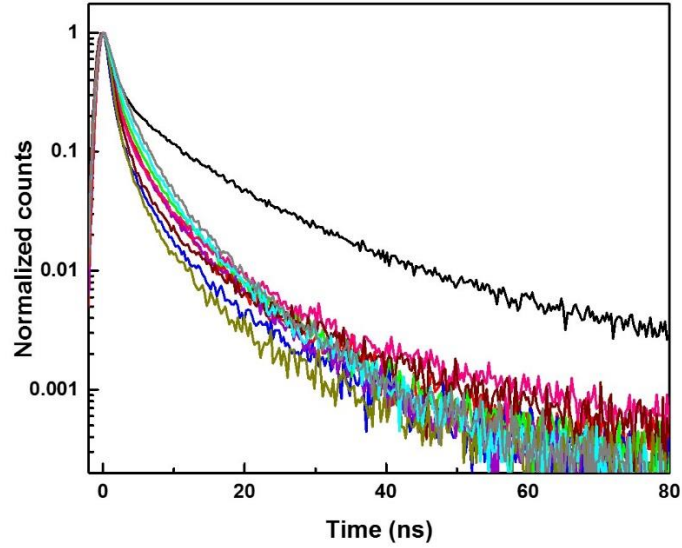

**Figure S1**: Fluorescence decay curves of a QD particle trapped inside the line defect (black) and a QD particle sandwiched between two GNRs forming a dimer structure based on multiple measurements on GNR-QD assemblies (colored). Black curve represent a typical decay curve for a QD particle without GNR.

|       | $\tau_1$ (ns) | $\tau_2$ (ns) |
|-------|---------------|---------------|
| QD    | 1.17          | 19.88         |
| QD-Au | 1.09          | 5.32          |
|       | 1.06          | 6.55          |
|       | 1.1           | 4.58          |
|       | 1.2           | 5.78          |
|       | 0.97          | 4.84          |
|       | 1.2           | 6.5           |
|       | 0.98          | 4.92          |
|       | 1.2           | 5.33          |
|       | 1.08          | 6.3           |
|       | 0.95          | 6.35          |

**Table S1**: The lifetime values extracted by fitting a double exponential equation to the fluorescence decay curves presented in Figure S3 above.

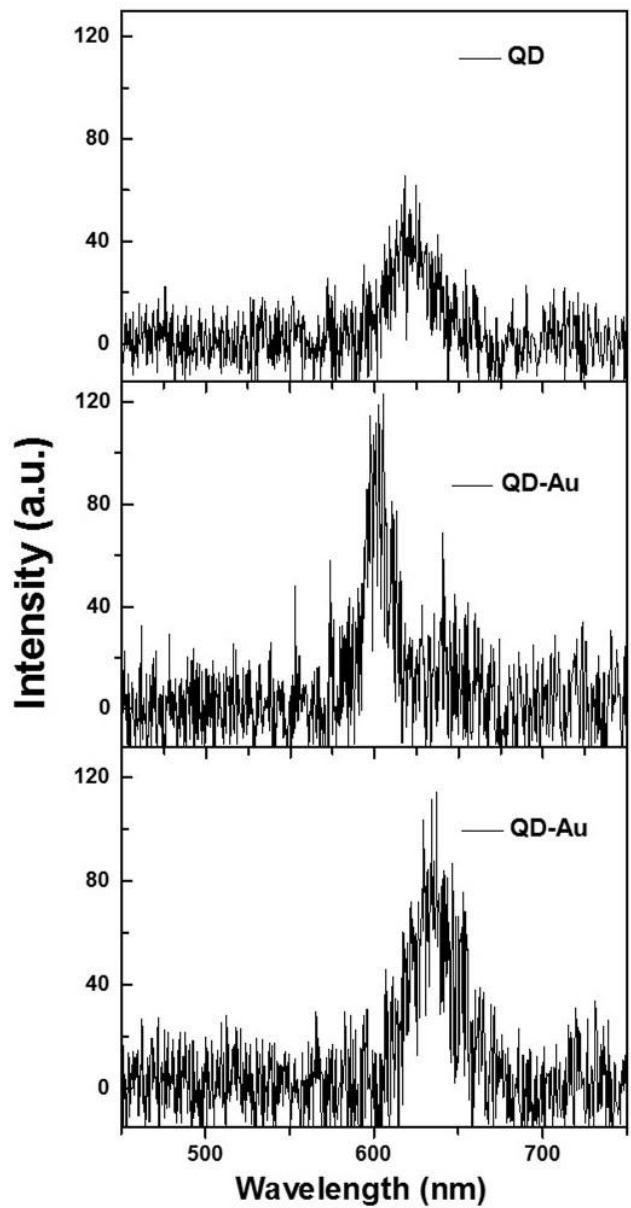

**Figure S2**: Fluorescence spectra of a single QD in line defect without GNR (a) and with GNRs showing a blueshifted (b) and redshifted (c) spectra.

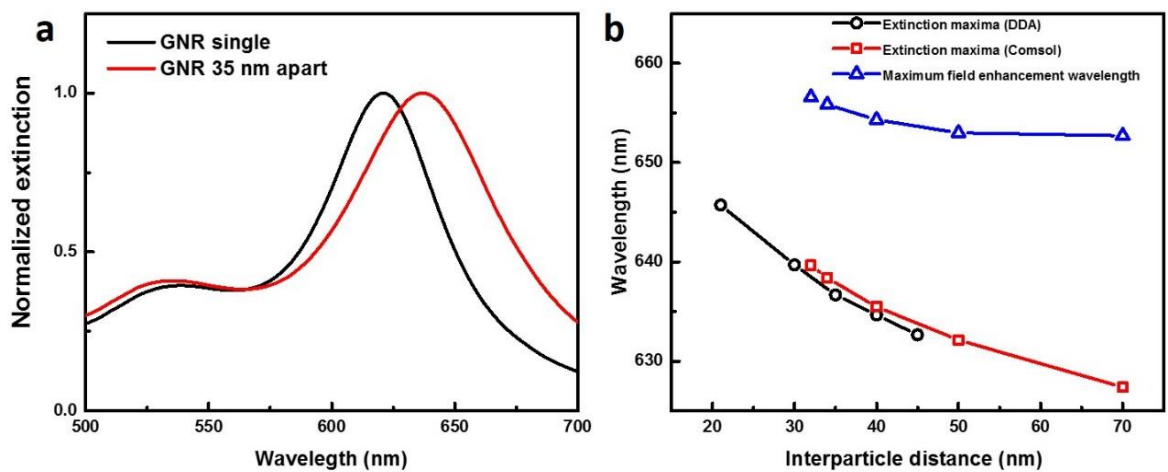

**Figure S3**: Simulated extinction spectra of a single GNR particle (black curve) and two GNR particles forming a sandwich structure (red curve) in the LC line defect. (b) Plot of LSPR peak positions vs. end-to-end separation between the gold cores of two GNRs located in the LC line defect calculated based on DDA (○) and COMSOL Multiphysics (□). Variations of maximum electric field enhancement wavelength (△) with end-to-end separation between the gold cores of two GNRs in the sandwich structure, calculated based on the electromagnetic simulations using COMSOL Multiphysics.